# Total angular momentum dichroism of the terahertz vortex beams at the antiferromagnetic resonances

A. A. Sirenko[1,2]*, P. Marsik[2], L. Bugnon[2], M. Soulier[2], C. Bernhard[2], T. N. Stanislavchuk[1], Xianghan Xu[3], and S.-W. Cheong[3]

[1] Department of Physics, New Jersey Institute of Technology, Newark, New Jersey 07102, USA

[2] Department of Physics, University of Fribourg, CH-1700 Fribourg, Switzerland

[3] Rutgers Center for Emergent Materials and Department of Physics and Astronomy, Rutgers University, Piscataway, New Jersey 08854, USA.

Terahertz vortex beams with different superposition of the orbital angular momentum $l=\pm1, \pm2, \pm3$, and $\pm4$ and spin angular momentum $\sigma=\pm1$ were used to study antiferromagnetic (AFM) resonances in TbFe$_3$(BO$_3$)$_4$ and Ni$_3$TeO$_6$ single crystals. In both materials we observed a strong vortex beam dichroism for the AFM resonances that are split in external magnetic field. The magnitude of the vortex dichroism is comparable to that for conventional circular dichroism due to $\sigma$. The selection rules at the AFM resonances are governed by the total angular momentum of the vortex beam: $j=\sigma+l$. In particular, for $l=\pm2, \pm3$, and $\pm4$ the sign of $l$ is shown to dominate over that for conventional circular polarization $\sigma$.

* Corresponding author. Email: sirenko@njit.edu

Magnetic properties of quantum matter are at the forefront of condensed matter physics. New optical probes are required for successful chiral discrimination in solids and for control and manipulation of quantum effects. By tapping the vorticity of the light beams and thereby adding an additional degree of freedom in a form of the orbital angular momentum (OAM) $l\hbar$ with $l = \pm 1, \pm 2, \pm 3,...$ [1], optics may facilitate future studies of topological phases [2], magnetic materials with non-collinear spins [3,4], and chiral excitations [5]. Vortex beams can simultaneously carry both OAM and conventional circular polarization in the form of the spin angular momentum (SAM) $\sigma\hbar$ with $\sigma = \pm 1$ [6]. Recently, broadband terahertz (THz) vortex beams have been implemented for spectroscopy of ferrimagnetic Dy iron garnet [7]. For beams with $l = \pm 1$, it has been demonstrated that the resonant absorption by magnetic modes of $Dy^{3+}$ depends on both the handedness of the vortex, or sign of $l$, and the direction of the beam propagation with respect to the sample magnetization. These experiments were complemented by theoretical studies of the optical activity of vortex beams [8]. Here we present qualitatively new experiments for the interaction between the spin waves, or magnons, with higher order vortex beams with OAM of $l = \pm 2, \pm 3, \pm 4$ in combination with a SAM of $\sigma = \pm 1$. Particularly, we will show that the selection rules for absorption of vortex beams at the antiferromagnetic (AFM) resonances in $TbFe_3(BO_3)_4$ and $Ni_3TeO_6$ are determined by the total angular momentum (AM) $j = l + \sigma$ of light. These two rather different AFM materials have been chosen to highlight the universal aspects of the observed dichroic effects for beams with OAM, SAM, and total AM.

The Ni-telluride (NTO) with the formula $Ni_3TeO_6$ and the Tb-ferroborate (Tb-FB) with the formula $TbFe_3(BO_3)_4$ have in common that the corresponding spins of the magnetic ions, $Fe^{3+}$ and $Ni^{2+}$, order collinearly along the crystalline $c$-axis below $T_N$ (Tb-FB) = 41 K [9,10] and $T_N$ (NTO) = 52 K [11]. Moreover, they have similar energies of the AFM resonances at $T << T_N$ of $\hbar\Omega_M$ (NTO) = 0.38 THz and $\hbar\Omega_M$ (Tb-FB) = 0.44 THz. In both materials $\hbar\Omega_M(T)$ decreases when $T \to T_N$. The crystal and magnetic structures of Tb-FB and NTO have several distinct features. $Ni_3TeO_6$ has a mono-polar and mono-chiral $R3$ structure with 10 atoms in a rhombohedral unit cell [12,13]. The $Ni^{2+}$ ions, which nominally have a $^3F_4$ configuration with $S=1$ and $L=3$, occupy three (i, ii, and iii) non-equivalent positions. Layers of $Ni^{2+}_iO_6/Te^{6+}O_6$ octahedra alternate along the $c$-axis with $Ni^{2+}_{ii}O_6/Ni^{2+}_{iii}O_6$ octahedra. The spin configuration of the $Ni^{2+}_i$-$Ni^{2+}_{ii}$-$N^{2+}_i$-$Ni^{2+}_{ii}$ network is up-up-down-down, while the $Ni^{2+}_{iii}$ spins are parallel to the $Ni^{2+}_{ii}$ spins in the same layer. More details about the spin and lattice excitations in NTO are given in Ref. [14]. Tb-FB, with two magnetic ions $Tb^{3+}$ and $Fe^{3+}$, is a multiferroic with a non-centrosymmetric structure at room temperature described by the trigonal space group $R32$ ($D_3^7$) that incorporates $FeO_6$ octahedra which form spiral chains along the *c*-axis [15]. At $T=200$ K, a structural phase transition takes place into a trigonal $P3_121$ phase [16]. Above 4 K, the AFM properties of Tb-FB are governed by the moments of the $Fe^{3+}$ ions in the $^6S_{5/2}$ configuration with $S=5/2$ and $L=0$. The $Tb^{3+}$ moments order only below 4 K and remain paramagnetic in all our experiments, albeit they are partially polarized via the *f–d* exchange interaction with the Fe moments [9,10,17]. The optical anisotropy and magnon spectra of Tb-FB are described in Refs. [18,19]. The phonons and crystal field transitions in ferroborates are discussed in Refs. [20,21].

The time-domain THz optical setup comprises photoconductive antennas as emitter and detector of coherent femtosecond pulses, a mirror optics with *f*-number 10, a liquid He flow optical cryostat, wire-grid linear polarizers, and an optical retarder [7,22]. External magnetic fields *B* of ±0.47 T were applied to the samples in the Faraday configuration using permanent ring magnets. The positive magnetic field $\vec{B}$ is parallel with the wave vector $\vec{k} \parallel \vec{z}$. The time-resolved signal was Fourier-transformed to obtain a continuous intensity spectrum $I(\nu)$ with frequencies $\nu$ between 0.1 THz and 1.65 THz. A single Fresnel prism made of TOPAS converted the incident linear polarization $(\vec{e}_x \pm \vec{e}_y)$ to the right-hand $\vec{e}_R = \vec{e}_x - i \cdot \vec{e}_y$ or left-hand $\vec{e}_L = \vec{e}_x + i \cdot \vec{e}_y$ circular polarizations. An axicon retarder made by TYDEX [23] was used to produce broad-band THz vortex beams with OAM $l=+1$ or $l=-1$ and with the electric field distribution of $\vec{e}_l(\vec{r},\phi) \approx (\vec{r}/r) \cdot \exp[i \cdot l \cdot (\phi - \phi_0)]$, where $\phi$ is the vortex beam phase, the initial phase is $\phi_0 = 3\pi/4$, and $\vec{r}$ is the radial coordinate [7]. To obtain an OAM with $|l|>1$, the broadband axicon was replaced with pairs of two identical transparent spiral plates that produced the Laguerre-Gaussian (LG) beams with the OAM at the sample with $\vec{e}_l(\phi) \approx \vec{e}_y \cdot \exp[i \cdot l \cdot \phi]$, where *l* is integer $l = \pm 2, \pm 3,$ or $\pm 4$ in the vicinity of the AFM resonances. The first spiral plate created the vortex beam at the sample, while the second plate unwound the vortex wave front at the detector. The handedness of the 3D printed spirals determines the sign of *l*. Several frequencies $\nu_l$ with integer *l*'s were simultaneously presented in the spectra $I(\nu)$ between 0.1 THz and 1.65 THz. The integer value of *l* at a given frequency $\nu_l$ (or wave length $\lambda_l$) is determined by the step height *h* and the refractive index *n*=1.56 of the spiral plate: $l \cdot \lambda_l = h(n-1)$ [24]. The spiral plates were produced with a Formlabs Form2 3D printer using clear V4 resin, similar to that in Ref.[25]. In addition, THz beams with a combined total AM of $j = l + \sigma$ were generated by transmitting circularly polarized beams with σ = ±1 through the same spiral plates producing LG beams with $\vec{e}_j(\phi) \approx \vec{e}_{L,R} \cdot \exp[i \cdot l \cdot \phi]$. See Supplemental Material for more details of the vortex beam characterization [26].

Figure 1 compares the magnetic dichroism for circularly polarized light with σ=±1 and for vortex beams with *l*=±1 at the AFM resonances of NTO at *T*=10 K and for *B*||*c* and shows that they have a similar magnitude and the same sign. As expected [27,28], the magnon peak splits into two modes at $\hbar\Omega_M^\pm(B) = \hbar\Omega_M(0) \pm \frac{1}{2} g_{Me} \mu_B B$, where $\mu_B$ is the Bohr magneton and $g_{Me}$ is the g-factor of the magnetic ion. For NTO we measured $g_{Ni} = 4.1 \pm 0.04$ and Ref.[19] reports that $g_{Fe}$ is close to 4. FIGs. 1(a,b) show the two modes of the magnon doublet $\hbar\Omega_M^\pm(B)$ that absorb light of the opposite circular polarization [27] and a reversal of *B* results in a switch of the selection rules for the sign of the circular polarization. FIGs. 1(c,d) show the corresponding absorption spectra for the vortex beams with OAM of *l*=±1 (and no circular polarization) as produced by the axicon retarder [7]. The two modes of the magnon doublet are equally sensitive to the sign of the OAM and the direction of *B* with respect to $\vec{k}$. A scheme of the circular and vortex dichroisms, which has the same sign for *l*, σ, and *j* is displayed in FIG. 1(e). For both NTO and Tb-FB, the vortex polarization of the magnon peaks is close to 100% and thus is much larger than the previously reported vortex dichroism of only ~22% in Dy iron garnet [7]. The strong dichroism in NTO and Tb-FB is due to the larger splitting of the magnon doublet of ~0.025 THz as compared to the linewidth γ of ~0.01 THz of the $\hbar\Omega_M^\pm(B)$ branches with $g_{Me}\mu_B B \gg \gamma$.

Next, we discuss the dichroism for the higher order vortex beams with OAM values of $l = \pm 2, \pm 3,$ and ±4 so that $\nu_{l=\pm 2,3,4}$ are in proximity with $\hbar\Omega_M^\pm(B,T)$ for *T* between 7 K and 30 K. Fig. 2(a,b) shows the magnon spectra of NTO and Tb-FB for the sample temperatures when $\hbar\Omega_M^\pm(B,T) \approx \nu_{l=\pm 3}$. FIG 2(c) shows the experimental configuration. Notably, the higher order vortex dichroism has the same sign and a similar magnitude as the one described above for $\sigma = \pm 1$ and $l = \pm 1$. An interesting difference between the vortex dichroism experiments with $l = \pm 1$ and $|l| \geq 2$ is as follows. The beams with $l = \pm 1$ [FIG. 1(c,d)] were produced with an axicon and consisted of a coherent combination of the radial and azimuthal modes [7]. Such a field has a non-zero curl of the electrical field $\vec{e}_{\pm 1}(\vec{r}, \phi)$ around the beam axis (see Supplemental Material [26]) that is equivalent to a pseudo magnetic field $\vec{B}_{pm} \approx \nabla \times \vec{e}_{\pm 1}(\vec{r}, \phi)$ directed along both the $\vec{k}$ vector and the AFM spin $\vec{S}$ ordering axis. The corresponding $\vec{S} \cdot \vec{B}_{pm}$ interaction may offer an intuitive way of understanding the vortex beam coupling to the ordered spins for experiments shown in FIG. 1(c,d). In contrast, the LG beams with $|l| \geq 2$ [FIG.2(a,b)] have a constant $\vec{e}_y$ component across the beam with $\vec{B}_{pm} \approx \nabla \times \vec{e}_y = 0$ but a nonzero $\vec{e}_z$ due to the vortex phase $\exp[i \cdot l \cdot \phi]$, where $\vec{e}_z$ is proportional to $|l|$. The coupling of such LG beams to the AFM spins may be similar to the spin-flip mechanism for intra Landau level transitions of 2D electrons in magnetic field [29]. Our experiment thus demonstrates that a coherent vortex phase and $\vec{e}_z \cdot \vec{S} \neq 0$ are sufficient to couple an LG vortex beam to the magnetic spin order in matter.

Thus far, we could not identify any strong magnetic modes, other than the magnetic dipoles at $\hbar \Omega_M^{\pm}(B,T)$, that might arise from higher-order spin resonances that are forbidden in the magnetic dipole approximation. However, a possible signature of such higher order resonances may be attributed to the asymmetry of the AFM modes shown in FIG. 2(b), which might arise from multi-spin excitations that involve the $Fe^{3+}$ and $Tb^{3+}$ spins in Tb-FB with the extra energy $\hbar \Omega_{Tb-Fe}(B,T)$. The shoulder on the low-frequency side in blue spectrum ($l$=+3) may correspond to $\hbar \Omega_M^{+}(B,T) - \hbar \Omega_{Tb-Fe}(B,T)$, while the high-frequency shoulder in the red spectrum ($l$=−3) may correspond to $\hbar \Omega_M^{-}(B,T) + \hbar \Omega_{Tb-Fe}(B,T)$. This interpretation is consistent with our findings that the shoulders are the strongest for integer values of $l$'s at $\hbar \Omega_M^{\pm}(B,T) \approx \nu_l$. The shoulders are absent in FIG. 2(a) for NTO which has only $Ni^{2+}$ magnetic ions.

To further study the interaction mechanism of the optical vortex with magnetism, we produced LG beams with combined orbital $l$ and spin $\sigma$ AM. FIGs 3(a,b) show spectra of Tb-FB measured at $T$'s so that $\hbar \Omega_M^{\pm}(B,T) \approx \nu_{l=\pm 3}$ for three $\sigma = 0, \pm 1$ and, correspondingly, $j = \pm 2, \pm 3, \pm 4$

. These spectra reveal that the AFM resonances coincide for a given $l = \pm 3$ for all three possible σ's demonstrating a strong total AM dichroism for which the selection rules of the magnon absorption are determined by the sign of *j*. Note that as long as |*j*|>>|σ|, the AFM resonances are no longer sensitive to the sign of σ. FIG. 4 shows that these selection rules apply not only for the integer values of *j* at $\hbar\Omega_M^\pm(B,T)$ (dark yellow arrows in FIGs 2, 3, and 4), but in the entire spectral range between |*j*|=2 and |*j*|=4. This suggests that the vortex beam at a non-integer multiple of the wavelength with respect to *h*, i.e. between *j* and *j*+1, behaves as a coherent superposition of the beams with *j* and *j*+1. Another interesting aspect concerns the conservation of the AM when light with a large total AM, e.g. with |*j*|=3, is absorbed by an AFM resonance with a magnetic dipole of $\Delta m_S = \pm 1$. This requires that the excess AM is transmitted to the crystal lattice, e.g. via the interaction with acoustic phonons.

The total AM dichroism in two AFM materials, NTO and Tb-FB, demonstrates that vortex beams with high orders of *l* can effectively couple to magnetic dipole excitations. The selection rules for the AFM resonances are dictated by the sign of *j* that dominates over that for conventional circular polarization given by σ. The different spin and orbital quantum numbers for the ground states of magnetic ions of $Ni^{2+}$ and $Fe^{3+}$ seem not to be important for the vortex beam absorption at the magnon frequencies. Thus, we predict a similar total AM dichroism for many AFM systems. The observed coupling between THz beams with high orders of the total AM with the ordered spin structure may provide an impetus for studies of non-local collective spin and carrier excitations in other quantum systems, such as Landau levels in 2D electron gas and graphene, chiral modes in Weyl and Dirac semimetals and *d*-wave superconductors.

The authors are grateful to K. Watt for the design of the spiral plates, to V. Kiryukhin for useful discussions, and to I. A. Gudim for the Tb-FB crystal. Work at the New Jersey Institute of Technology and Rutgers was supported by the U.S. DOE under Contract No. DEFG02-07ER46382. The work at the University of Fribourg was funded by the Schweizerische Nationalfonds (SNF) by Grant No. 200020-172611. AAS is grateful to SNF for the IZSEZ0-184564 support during his research visit at the University of Fribourg.

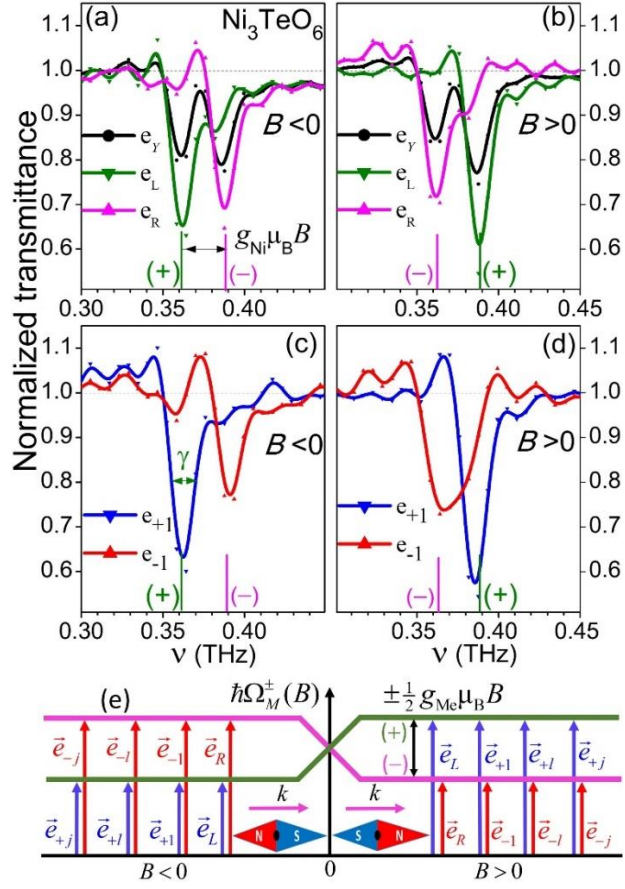

FIG. 1 (color online) Magnetic dichroism for circularly polarized light and vortex beams with $l=\pm1$ in NTO. (a,b) Normalized transmittance for circularly polarized light, $\vec{e}_L$ and $\vec{e}_R$, and for conventional linear polarization $\vec{e}_y$. (c,d) Normalized transmittance for broadband vortex beams with $\vec{e}_{+1}$ (blue spectra) and $\vec{e}_{-1}$ (red spectra). All experimental data in (a-d) are taken at $B=\pm0.47$ T and $T=10$ K and are normalized to that measured at $T=60$ K in the same field: $I(\nu,T)/I(\nu,T=60\text{ K})$. (e) Schematics of the experimentally determined selection rules for the AFM resonances using both, circularly polarized light and vortex beams with $l$ and $j$. The positive $\vec{B}$ is parallel with $\vec{k}$.

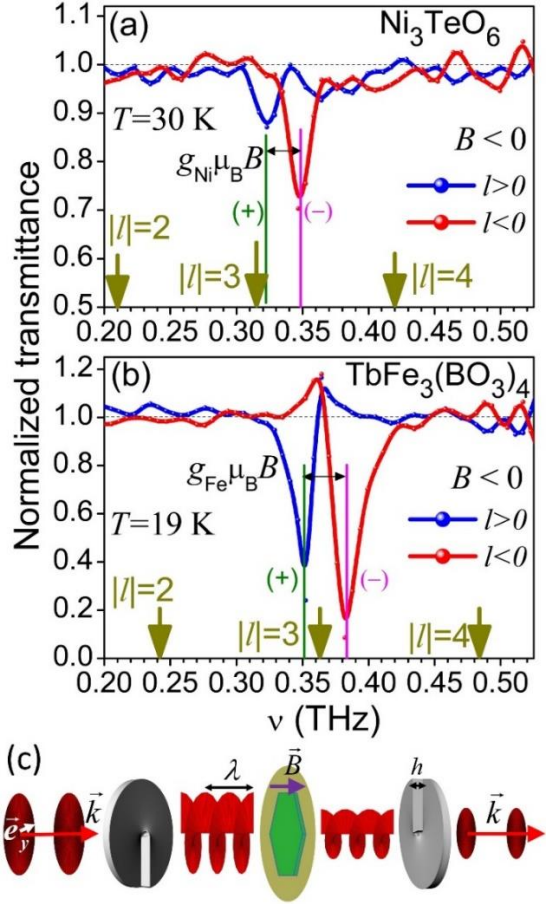

FIG. 2 (color online) Vortex dichroism for higher orders of the OAM for (a) NTO at $T$=30 K and (b) Tb-FB at $T$=19 K. Normalized transmittance spectra for two linearly polarized LG vortex beams $\vec{e}_{l>0}$ (blue spectra) and $\vec{e}_{l<0}$ (red spectra) were taken at $B = -0.47$ T that brings the AFM resonances close to $\nu_{l=\pm 3}$. The corresponding AFM resonances $\hbar\Omega_M^\pm(B,T)$ are marked with vertical olive and magenta lines. The frequencies that correspond to $\nu_{l=\pm 2,\pm 3,\pm 4}$ are marked with bold dark yellow arrows. Note the absorption peak asymmetry for the $\vec{e}_{l=+3}$ and $\vec{e}_{l=-3}$ AFM resonances in (b). (c) Schematics for the measurement geometry with linear polarization $\vec{e}_y$ for the THz beam input. Sample (green hexagon) is placed between a pair of identical spiral plates with step height $h$.

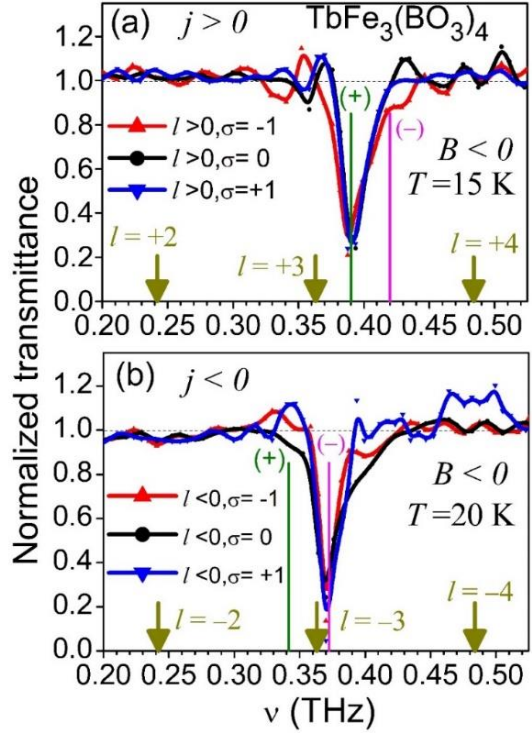

FIG. 3 (color online) Normalized transmittance spectra for the LG vortex beams for (a) $j>0$ and (b) $j<0$ in Tb-FB at $B=-0.47$ T. The resonance frequencies for $\nu_{l=\pm 2}$, $\nu_{l=\pm 3}$, and $\nu_{l=\pm 4}$ are marked with bold dark yellow arrows. The temperatures were chosen to bring the AFM resonances $\hbar\Omega_M^\pm(B,T)$ shown with vertical olive and magenta lines close to $\nu_{l=\pm 3}$ in both, (a) and (b). Note that the corresponding AFM resonances coincide for $j=+2$ $(l=+3,\sigma=-1)$, $j=+3$ $(l=+3,\sigma=0)$, and $j=+4$ $(l=+3,\sigma=+1)$ in (a) and for $j=-2$ $(l=-3,\sigma=+1)$, $j=-3$ $(l=-3,\sigma=0)$, and $j=-4$ $(l=-3,\sigma=-1)$ in (b). The measurement geometry is the same as in FIG. 2(c) but with various input polarizations $\vec{e}_y$ for black, $\vec{e}_L$ for blue, and $\vec{e}_R$ for red spectra.

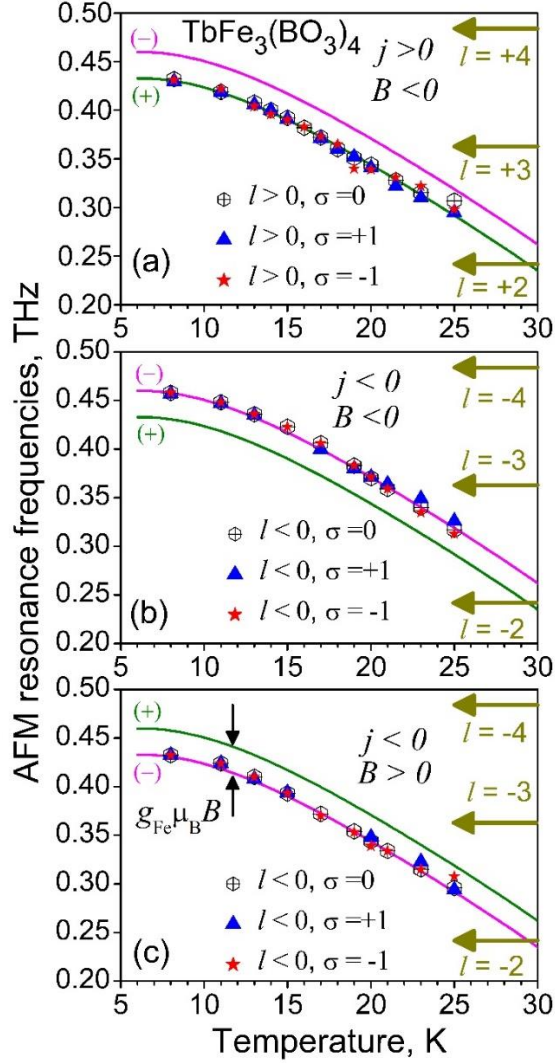

FIG. 4. (color online) Total angular momentum dichroism in Tb-FB revealed in the temperature dependencies of the AFM resonance frequencies $\hbar\Omega_M^\pm(B,T)$ for different combinations of the $\vec{B}$ direction and the sign of $j$ for the vortex beams. (a) $j > 0$, $B < 0$, (b) $j < 0$, $B < 0$, and (c) $j < 0$, $B > 0$. Solid curves correspond to the AFM frequencies $\hbar\Omega_M^\pm(B,T)$ measured using conventional circularly polarized light. The color code for the curves (olive and magenta) is the same as for $\hbar\Omega_M^\pm(B)$ lines in FIG. 1(e). The measurement geometry is the same as shown in FIG. 2(c) but with various input polarizations $\vec{e}_y$ for black, $\vec{e}_L$ for blue, and $\vec{e}_R$ for red symbols.

[1] L. Allen, M. W. Beijersbergen, R. J. C. Spreeuw, and J. P. Woerdman, Phys. Rev. A **45**, 8185-8189 (1992).
[2] M. Z. Hasan and C. L. Kane, Rev. Mod. Phys. **82**, 3045 (2010).
[3] S.-W. Cheong and M. Mostovoy, Nature Mater **6**, 13–20 (2007).
[4] S.-W. Cheong, D. Talbayev, V. Kiryukhin, and A. Saxena, npj Quant Mater **3**, 19 (2018).
[5] W. Koshibae and N. Nagaosa, Nature Communications **7**, 10542 (2016).
[6] C. T. Schmiegelow, J. Schulz, H. Kaufmann, T. Ruster, U. G. Poschinger & F. Schmidt-Kaler, Nature Communications **7**, 12998 (2016).
[7] A. A. Sirenko, P. Marsik, C. Bernhard, T. N. Stanislavchuk, V. Kiryukhin, and S-W. Cheong, Phys. Rev. Lett. **122**, 237401 (2019).
[8] K. A. Forbes and D. L. Andrews, Phys. Rev. Research **1**, 033080 (2019).
[9] C. Ritter, A. Balaev, A. Vorotynov, G. Petrakovskii, D. Velikanov, V. Temerov, and I. Gudim, J. Phys.: Condens. Matter **19**, 196227 (2007).
[10] M.N. Popova, T.N. Stanislavchuk, B.Z. Malkin, L.N. Bezmaternykh, J. Phys.: Condens. Matter **24**, 196002 (2012).
[11] I. Zivkovic, K. Prsa, O. Zaharko, and H. Berger, J. Phys.: Condens. Matter **22**, 056002 (2010).
[12] R. E. Newnham, and E. P. Meagher, Mater. Res. Bull. **2**, 549-554 (1967).
[13] R. Becker, and H. Berger, Acta Cryst. E**62**, i222-i223 (2006).
[14] S. Skiadopoulou, F. Borodavka, C. Kadlec, F. Kadlec, M. Retuerto, Zh. Deng, M. Greenblatt, and S. Kamba, Phys. Rev. B **95**, 184435 (2017).
[15] J. C. Joubert, W. B. White, and R. J. Roy, J. Appl. Cryst.**1**, 318 (1968).
[16] D. Fausti, A. A. Nugroho, P. H. M. van Loosdrecht, S. A. Klimin, M. N. Popova, and L. N. Bezmaternykh, Phys. Rev. B **74**, 024403 (2006).
[17] M. N. Popova, E. P. Chukalina, T. N. Stanislavchuk, and L. N. Bezmaternykh, J. Magn. Magn. Mater. **300**, e440 (2006).
[18] A. B. Kuz'menko, A. A. Mukhin, V. Yu. Ivanov, A. M. Kadomtseva, S. P. Lebedev, and L. N. Bezmaternykh, J. Exp. Theor. Phys. **113**, 113–120, (2011), [first published in Russian in Zhurnal Eksperimental'noi i Teoreticheskoi Fiziki **140**, pp. 131–139, (2011)].
[19] D. Szaller, V. Kocsis, S. Bordács, T. Fehér, T. Rõõm, U. Nagel, H. Engelkamp, K. Ohgushi, and I. Kézsmárki, Phys. Rev. B **95**, 024427 (2017).
[20] M. N. Popova, K. N. Boldyrev, S. A. Klimin, T. N. Stanislavchuk, A. A. Sirenko, L. N. Bezmaternykh, Journal of Magnetism and Magnetic Materials **383**, 250–254 (2015).
[21] K. N. Boldyrev, T. N. Stanislavchuk, A. A. Sirenko, D. Kamenskyi, L. N. Bezmaternykh, M. N. Popova, Phys. Rev. Lett. **118**, 167203 (2017).
[22] P. Marsik, K. Sen, J. Khmaladze, M. Yazdi-Rizi, B. P. P.Mallett, and C. Bernhard, Appl. Phys. Lett. **108**, 052901 (2016).
[23] http://www.tydexoptics.com/products/thz_optics/thz_lens/#thz_axi
[24] J. Sun, J. Zeng, X. Wang, A. N. Cartwright, and N. M. Litchinitser, Scientific Reports **4**, 4093 (2014).
[25] Changming Liu, Jinsong Liu, Liting Niu, Xuli Wei, Kejia Wang, and Zhengang Yang Scientific Reports **7**, 3891 (2017).

[26]  See Supplemental Material at *** for details of experimental setup.
[27]  F. Kepfer and C. Kittel, Phys. Rev. **85**, 329-337 (1952).
[28] T. Oguchi and A. Honma, J. Phys. Soc. Jpn. **16**, 79-94 (1961).
[29]  E. I. Rashba and V. I. Sheka, Modern Problems in Condensed Matter Sciences **27**, 131–206, (1991).

SUPPLEMENTAL MATERIALS for

**Total angular momentum dichroism of the terahertz vortex beams at the antiferromagnetic resonances**

A. A. Sirenko[1,2]*, P. Marsik[2], L. Bugnon[2], M. Soulier[2], C. Bernhard[2], T. N. Stanislavchuk[1], Xianghan Xu[3], and S.-W. Cheong[3]

[1] Department of Physics, New Jersey Institute of Technology, Newark, New Jersey 07102, USA

[2] Department of Physics, University of Fribourg, CH-1700 Fribourg, Switzerland

[3] Rutgers Center for Emergent Materials and Department of Physics and Astronomy, Rutgers University, Piscataway, New Jersey 08854, USA.

* Correspondence and requests for materials should be addressed to A. A. S. (email: sirenko@njit.edu )

1. **Experimental setups for the vortex beams**

Fig. S1 shows schematics for the axicon-based experimental setup used for the generation of the vortex beams in FIG. 1(c,d) of the manuscript. The setup has been previously described in Ref. [1]. It consists of a linear polarizer, a two-bounce Fresnel prism (FP) made of TOPAS, and a 4-bounce axicon made of transparent silicon. The two-bounce FP served as a broad-band retarder to convert the input linear polarization $(\vec{e}_x \pm \vec{e}_y)$ to the right-hand $\vec{e}_R = \vec{e}_x - i \cdot \vec{e}_y$ or left-hand $\vec{e}_L = \vec{e}_x + i \cdot \vec{e}_y$ circular polarizations. An axicon retarder is a converter between circularly polarized light and vortex beams, or between light with spin angular momentum (SAM) and orbital angular momentum (OAM). It was designed to produce broad-band THz vortex beams with OAM $|l|=1$. The electric field distribution in the output beam is $\vec{e}_l(\vec{r},\phi) \approx (\vec{r}/r) \cdot \exp[i \cdot l \cdot (\phi - \phi_0)]$, where $\phi$ is the vortex phase, $l=+1$ or $l=-1$, the initial phase is $\phi_0 = 3\pi/4$, and $\vec{r}$ is the radial coordinate. The switch between $l=+1$ and $l=-1$ outputs is produced by rotation of the linear polarizer at the input of the setup between two orthogonal positions: $(\vec{e}_x + \vec{e}_y)$ and $(\vec{e}_x - \vec{e}_y)$. Calculated radially independent electric fields $\vec{e}_{+1}(\vec{r},\phi)$ and $\vec{e}_{-1}(\vec{r},\phi)$ are also shown in Fig. S1. Note that such beams have a non-zero curl of the electrical field $\vec{e}_{\pm 1}(\vec{r},\phi)$ around the beam axis $\nabla \times \vec{e}_{\pm 1}(\vec{r},\phi) \neq 0$ that is collinear with the $\vec{k}$ vector of the light. Note here that the circularly polarized light at the input of the axicon cannot "leak" through. First, the axicon uses 4-bounces of the total internal reflections, the same as in a 4-bounce Fresnel prism that is known to be a perfect converter between circular and linear polarizations. Furthermore, because time-domain spectroscopy is used, any "leaking" circularly polarized photons that bypass the axicon or those not experiencing the expected four bounces would arrive at the detector at a completely different time than the major pulse. Thus, the leaking photons cannot contribute to the measured interferogram.

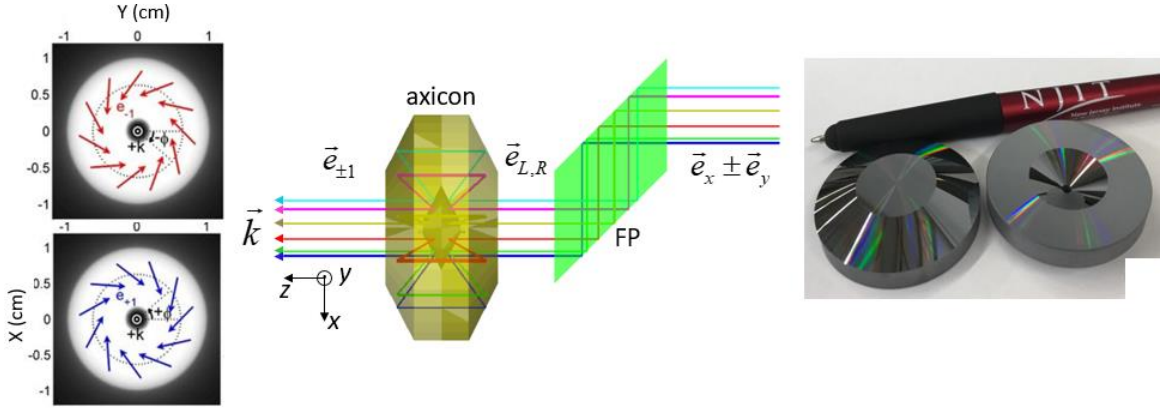

FIG. S1. The central panel shows the conversion of the circularly polarized light into vortex beam modes $\vec{e}_{+1}(\vec{r},\phi)$ and $\vec{e}_{-1}(\vec{r},\phi)$ using a combination of a two-bounce Fresnel prism (FP) and an axicon (adapted from Ref. [1]). After passing through an FP retarder, the linearly polarized light $(\vec{e}_x + \vec{e}_y)$ or $(\vec{e}_x - \vec{e}_y)$ becomes circularly polarized $\vec{e}_L = \vec{e}_x + i \cdot \vec{e}_y$ or $\vec{e}_R = \vec{e}_x - i \cdot \vec{e}_y$. After passing through the axicon, the beam acquires a vortex phase $\vec{e}_{+1}(\vec{r},\phi)$ or $\vec{e}_{-1}(\vec{r},\phi)$ while losing its circular polarization. The electric field distribution is shown in the left panel where the $\vec{k}$ vector is normal to the page. The right panel shows a photo of the axicon retarder.

Fig. S2(a,b) shows schematics for the experimental setups with the spiral plate used for the generation of the Laguerre-Gaussian (LG) vortex beams with $|l|>1$. These setups were used for experimental data in FIG. 2, FIG. 3, and FIG. 4 of the manuscript. To obtain an OAM with $|l|>1$, the broadband axicon was replaced with pairs of two identical 3D-printed transparent spiral plates. The spiral plates were produced with a Formlabs Form2 3D printer using clear V4 resin. In FIG. S2(a) the polarization input is linear $\vec{e}_y$ that produced the OAM at the sample position with the field distribution of $\vec{e}_l(\phi) \approx \vec{e}_y \cdot \exp[i \cdot l \cdot \phi]$, where $l$ takes integer values of $l = \pm 2, \pm 3,$ or $\pm 4$ at the corresponding frequencies $\nu_l$. The field distribution across the beam at the sample position is shown in the right panel. Note that the first spiral plate created the vortex beam at the sample, while the second plate unwound the vortex wave front such that a plane wave propagates further to the detector. The handedness of the 3D printed spirals determines the sign of $l$. Several frequencies $\nu_l$ with integer values of $l$'s were simultaneously present in the measured spectra $I(\nu)$. The integer value of $l$ at a given frequency $\nu_l$ (or wavelength $\lambda_l$) is determined by the step height $h$ and the refractive index $n=1.56$ of the spiral plate: $l \cdot \lambda_l = h(n-1)$ [2]. FIG. S2(b) shows the same setup but with a circularly polarized input polarization. Such THz beams have a combined total angular momentum (AM) of $j = l + \sigma$ with $\sigma = \pm 1$ and integer values of $l = \pm 2, \pm 3,$ or $\pm 4$ and the field distribution $\vec{e}_j(\phi) \approx \vec{e}_{L,R} \cdot \exp[i \cdot l \cdot \phi]$ across the LG vortex beams. Both beams shown in FIG.S2(a,b) have a constant field across the beam with no curl: $\nabla \times \vec{e}_y = 0$ and $\nabla \times \vec{e}_{R,L} = 0$.

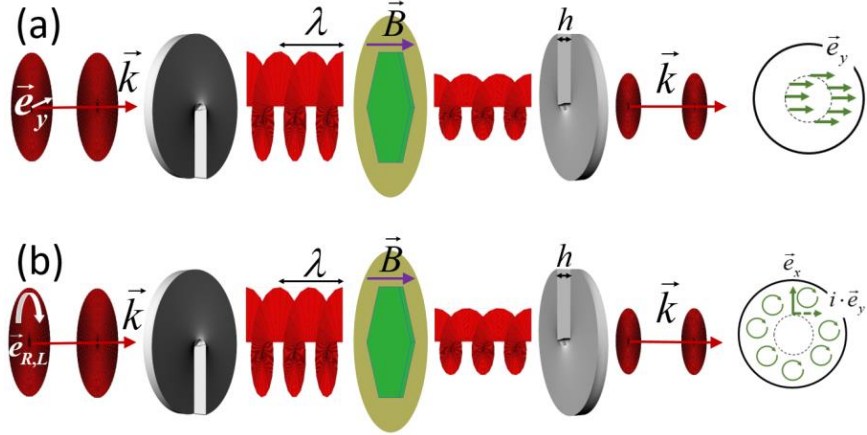

FIG. S2 (a) Schematics for the measurement geometry with a linear polarization $\vec{e}_y$ for the THz beam input. (b) The same with a circular polarization $\vec{e}_{R,L}$ for the THz beam input. Sample (green hexagon) is placed between a pair of identical spiral plates with step height $h$.

## 2. Experimental characterization of the THz vortex beams

For experimental characterization of the THz vortex beams produced with the 3D printed spiral plates we used the same experimental setup as for the sample measurements. We mounted an empty metal aperture at the sample position on an XY motorized translational stage. The distance between the aperture and the spiral plate was 11 cm. The optical setup consisted of a rotatable linear polarizer, a Fresnel two-bounce prism, a vortex plate that produced OAM of $l=2$ at the light frequency of 0.5 THz ($\lambda= 0.6$ mm), and an aperture (see FIG. S3). The beam footprint at the sample position has the measured full width at half maximum (FWHM) of about 9 mm. The aperture diameter was chosen to be 2 mm, which allowed us to keep the transmitted signal at a high signal-to-noise ratio. The cross section of the beam was scanned with the steps $\Delta x$ and $\Delta y$ of 1 mm, or half the aperture diameter. This beam characterization has two goals: (i) to demonstrate the expected toroidal intensity shape and the phase variation around the beam, and (ii) to demonstrate that the linear and circular polarizations of the photons remains nearly unchanged after passing through the spiral plate. In the following we will demonstrate that the vortex beams have the expected shape and phase, and they can also carry spin angular momenta, also known as right- and left-hand circular polarizations.

FIG. S4 shows a map for the beam amplitude, two cross sections along X and Y directions and an integrated radial distribution of the amplitude. The input linear polarizer is at zero angle P with respect to the $y$-axis (see FIG.S1 for x-y-z notation), producing $\vec{e}_y$. After passing through the Fresnel prism, the beam polarization does not change. As expected, the beam has a toroidal shape with the Amplitude minimum close to the center. The outside dimensions of the beam with FWHM of about 8 mm are determined by the focusing optics for the original Gaussian beam produced by the 50 mm-diameter parabolic mirror with the apparent focus distance of 500 mm. The measured amplitude in the very center of the beam is at 20% of the maximum that is very reasonable for the measurement with an aperture whose dimensions are close to the diffraction limit. Note that the same diffraction effect results in the beam broadening at the sample position. In the actual experiments we estimated the FWHM of the unrestricted beam to be closer to 7 mm at the sample. By rotating the input linear polarizer $P$ to +45° or +315° (=-45°) , we produced one of the two circular polarizations at the vortex spiral plate with $\vec{e}_R = \vec{e}_x - i \cdot \vec{e}_y$ or $\vec{e}_L = \vec{e}_x + i \cdot \vec{e}_y$, respectively. Still, the corresponding intensity maps did not change significantly and they are very similar to that in FIG. S4. The Amplitude maps and cross sections for $P = +45°$ and +315° are shown in FIG. 5.

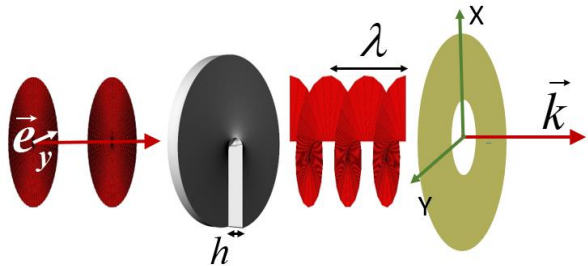

FIG.S3. Schematics of the experimental setup for the vortex beam characterization. The aperture diameter is 2 mm, the distance between the spiral plate and the aperture is 11 cm. The input linear polarization $\vec{e}_y$ (or circular polarizations $\vec{e}_R = \vec{e}_x - i \cdot \vec{e}_y$ and $\vec{e}_L = \vec{e}_x + i \cdot \vec{e}_y$) are produced with a combination of a linear polarizer and a Fresnel prism.

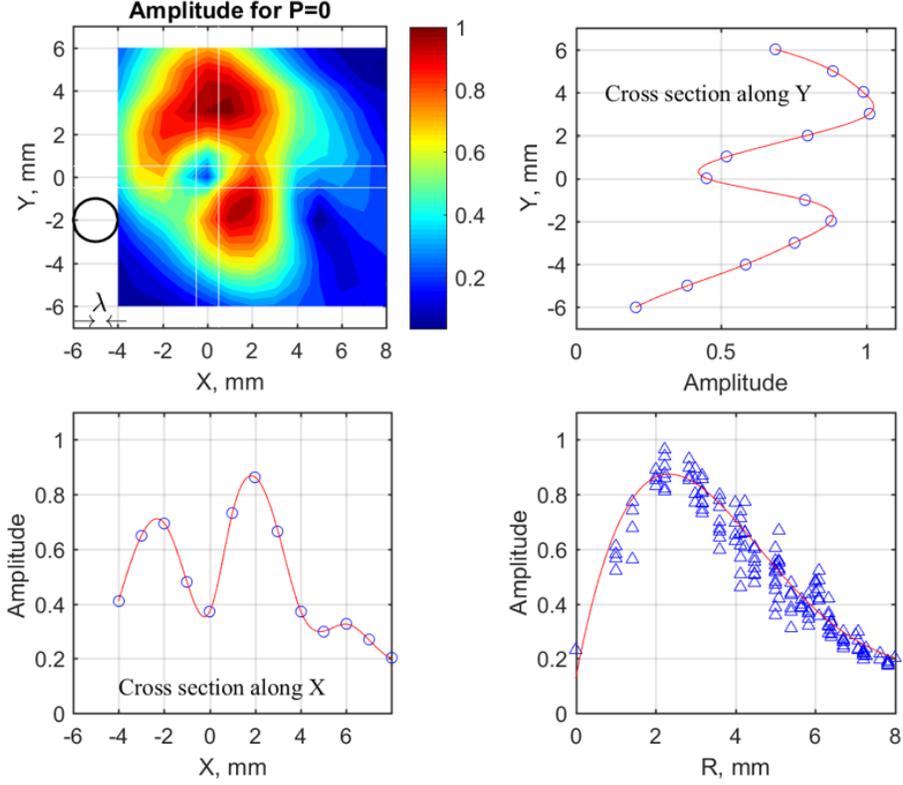

FIG.S4. The normalized 12 mm × 12 mm map of the signal Amplitude at 0.5 THz for the input linear polarization with P = 0 ($\vec{e}_y$). Black circle shows the aperture size; the wavelength is shown with two black arrows. The side panels present the amplitude cross sections along the bands shown with white lines. The radial dependence of the signal amplitude shows the intensity peak at about 3 mm with respect to the center of the beam and a minimum of amplitude at the center. Red curves guide the eye.

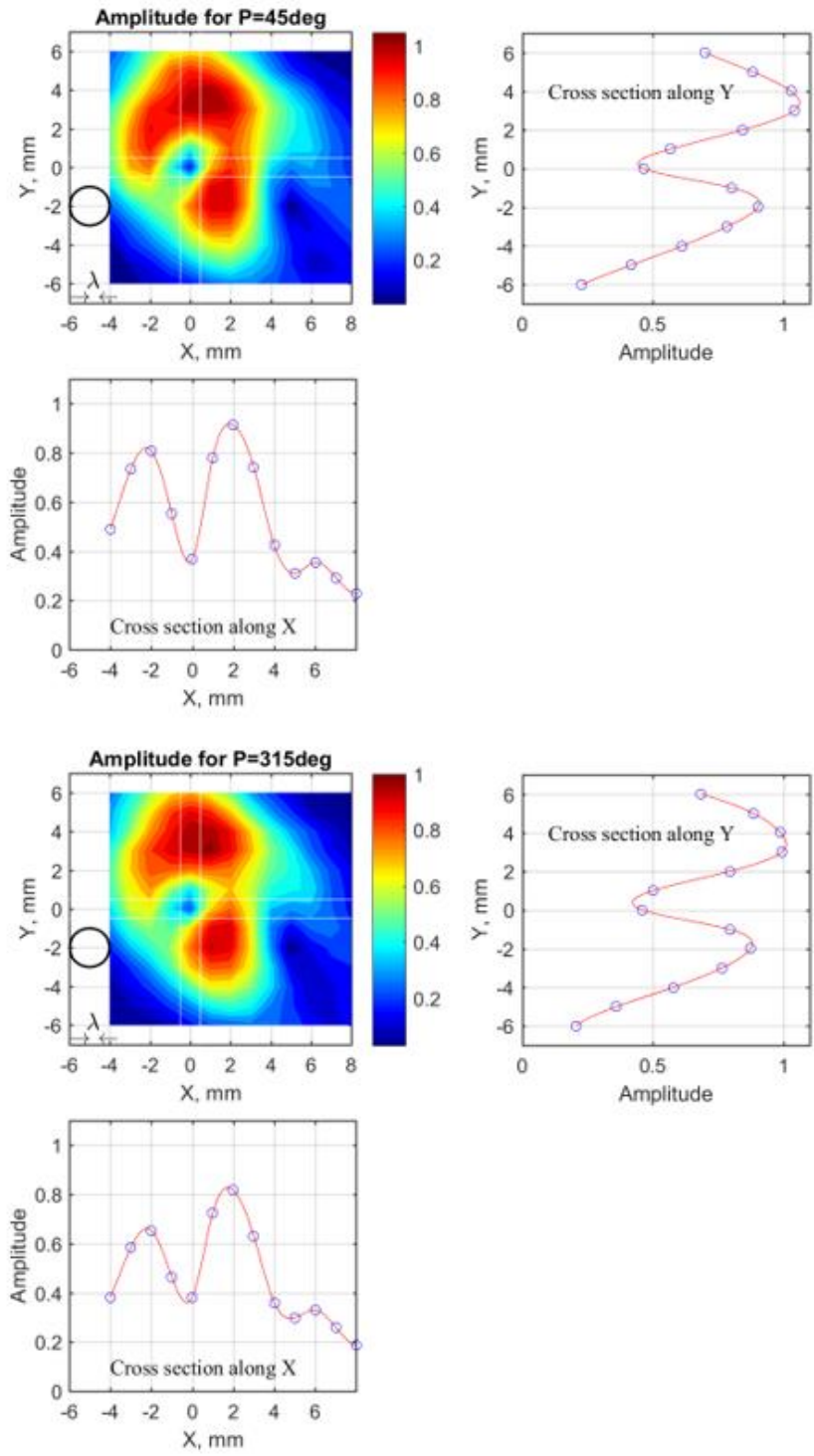

FIG.S5. The same as in FIG. S4 for the two circularly polarized inputs with P = 45° (top) and P = 315° (bottom).

FIG. S6 shows phase maps measured for three input linear polarizations $P=0$, $+45°$, and $+315°$. In addition, the same Figure shows azimuthal variation of the map for 360 degrees around the beam propagation direction at $R=3$ mm. As expected, the phase for the beam produced with a spiral plate with $l=2$ changes by $2\pi l$ for 360° rotation around the beam direction. The bottom panels show the experimental points from the map and red solid lines for theoretical expectation for the beam phase. Two very similar dependencies for the phase variation were measured for $P=+45°$, and $+315°$ as well (see FIG. S6).

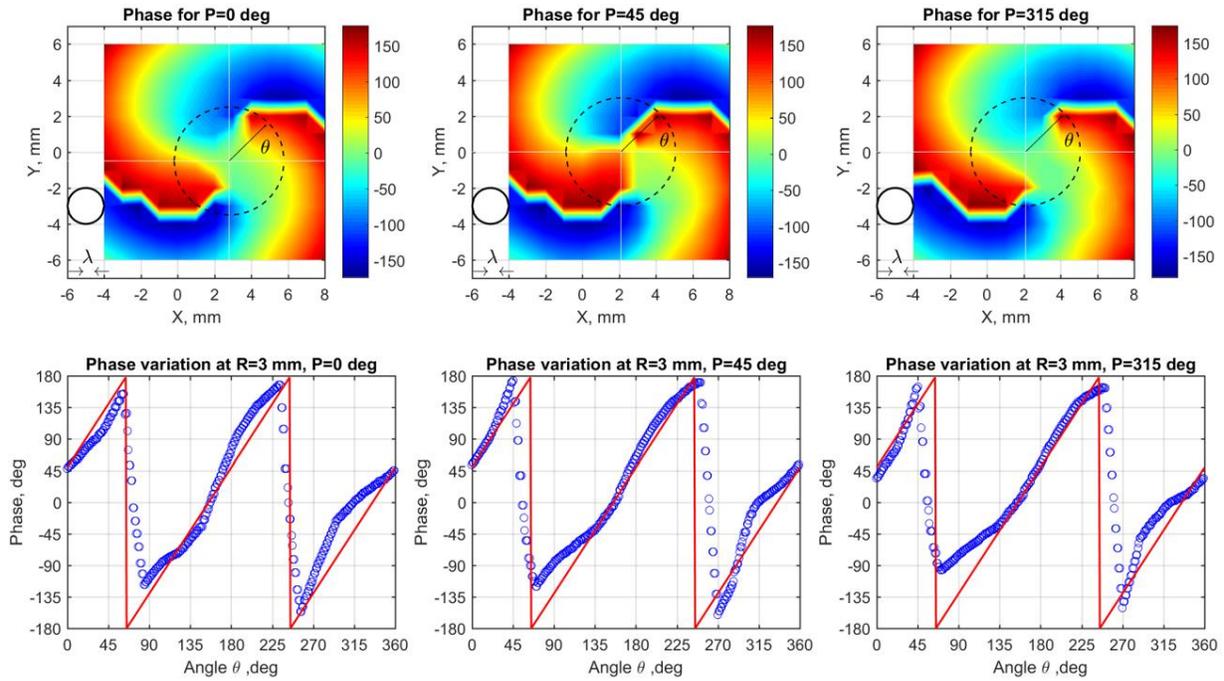

FIG.S6. The signal phase at 0.5 THz for the input linear polarization with $P = 0$ (left), $P = 45°$ (center), and $P = 315°$ (right). (Black circle shows the aperture size; the wavelength is shown with two black arrows. The bottom panels present the azimuthal dependence of the phase extracted from the map along the black dotted circle with $R=3$ mm. Red lines show theoretical dependence of the phase that changes by $4\pi$ for the vortex plate with $l=2$ at 0.5 THz.

Let's turn our attention to the polarization properties of the individual rays composing the vortex beams. In the first approximation, the polarization state of individual rays passing through the spiral plate should be the same as created by the combination of the linear polarizer and the Fresnel prism. The only deviations from that could have been produced by imperfections of the spiral plates, that however were printed in a solid form with 0.025 mm resolution that is ~ 25 times better than the wavelength $\lambda$= 0.6 mm. For experimental characterization of the beam polarization we used the conventional rotating analyzer method [3]. The experimental results are presented below using the following Jones vector conventions. The ratio of the electric field amplitudes $|\vec{e}_x/\vec{e}_y| = \tan\Psi$ and the phase difference between the two complex vectors $\vec{e}_y$ and $\vec{e}_x$ is $\Delta$. In this notation, light with linear polarization along y, should have $\Psi = 0$ and undefined $\Delta$. For two circular polarizations, $\vec{e}_L = \vec{e}_x + i\cdot\vec{e}_y$ and $\vec{e}_R = \vec{e}_x - i\cdot\vec{e}_y$, the following values are expected: $\Psi = 45°$ and $\Delta = \pm 90°$.

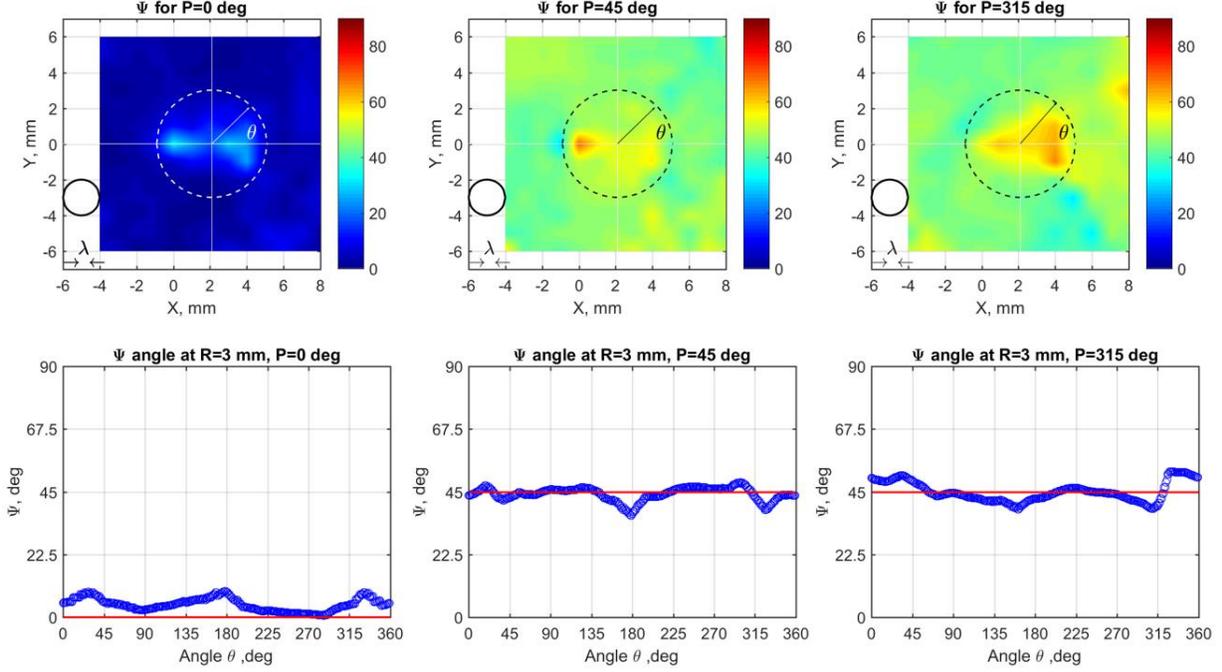

FIG.S7. The polarization angles $\Psi$ at 0.5 THz for the input linear polarization with P = 0 (left), P = 45° (center), and P = 315° (right). The bottom panels present the azimuthal dependence of the $\Psi$ values extracted from the map along the black dotted circle with R=3 mm. Red lines show theoretical value of $\Psi$ being 0 for linear polarization and 45° for both circular polarizations.

FIG. S7 and FIG. 8 show experimental maps for $\Psi$ and $\Delta$, correspondingly, measured for *P*=0, +45°, and +315°. The azimuthal variations are shown with blue symbols along with the corresponding theoretical expectations shown with horizontal red lines. The close proximity between the experiment and theory provides a proof that the THz vortex beams used in our experiments possess simultaneously the vortex phase, which is the property of the beam, and the spin angular momentum $\sigma = \pm 1$ for the corresponding input circular polarizations and $\sigma = 0$ for the input with a linear polarization. Our results for intensity maps are similar to that previously reported for 3D spiral plates used to produce THz vortex beams [4]. Note that the minor deviation of our experimental data from the theory may be primarily due to the small aperture diameter of 2 mm that brings the beam mapping close to the diffraction limit for λ= 0.6 mm. We believe that the unrestricted vortex beams are closer to the theoretical expectation than the mapping results that are presented here.

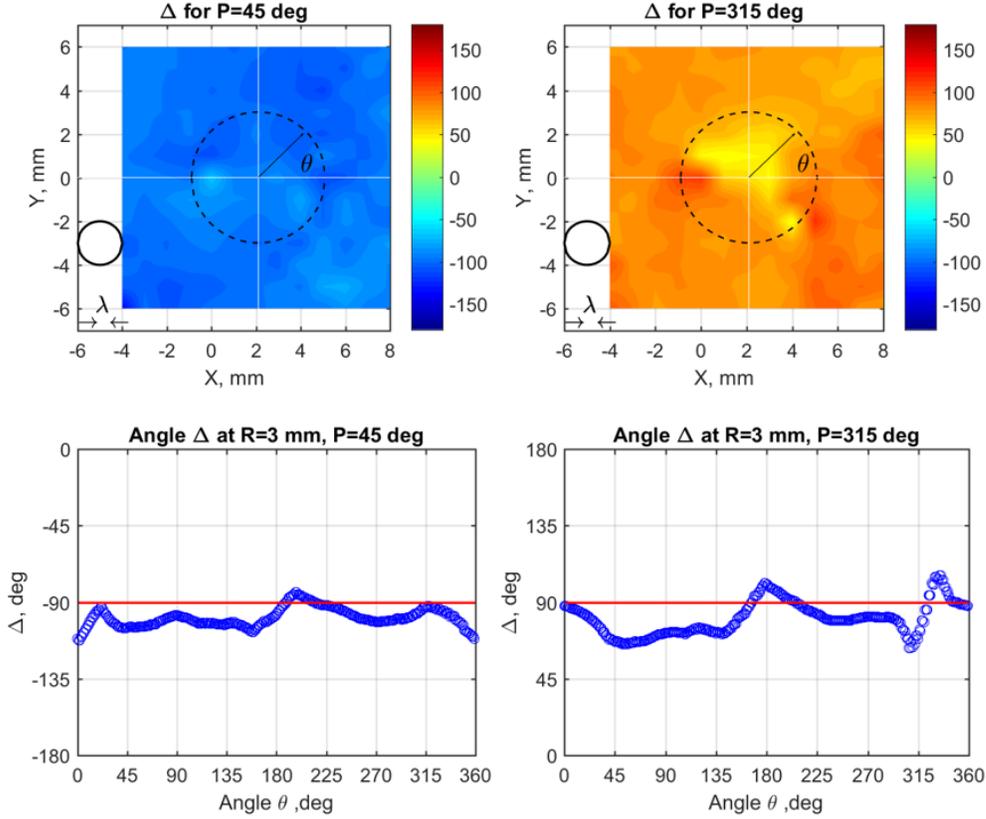

FIG.S8. The polarization angles Δ at 0.5 THz for inputs with two circular polarizations with P = 45° (left) and P = 315° (right). The bottom panels present the azimuthal dependence of the Δ values extracted from the map along the black dotted circle with R=3 mm. Red lines show the theoretical value of Δ being +90° and -90° for two circular polarizations. Note that Δ is undefined for the linear polarization input with P = 0 and, thus, it is not shown here.

[1] A. A. Sirenko, P. Marsik, C. Bernhard, T. N. Stanislavchuk, V. Kiryukhin, and S-W. Cheong, Phys. Rev. Lett. **122**, 237401 (2019).

[2] J. Sun, J. Zeng, X. Wang, A. N. Cartwright, and N. M. Litchinitser, Scientific Reports **4**, 4093 (2014).

[3]     P. Marsik, K. Sen, J. Khmaladze, M. Yazdi-Rizi, B. P. P.Mallett, and C. Bernhard, Appl. Phys. Lett. **108**, 052901 (2016).

[4] Changming Liu, Jinsong Liu, Liting Niu, Xuli Wei, Kejia Wang, and Zhengang Yang Scientific Reports **7**, 3891 (2017).